\documentclass{ws-ijgmmp}

\begin{document}

\markboth{P.K. Sahoo, Parbati Sahoo, Binaya K. Bishi }
{Anisotropic cosmological models in $f(R,T)$ gravity with variable deceleration parameter}

%
\catchline{}{}{}{}{}
%

\title{\textbf{ANISOTROPIC COSMOLOGICAL MODELS IN $f(R,T)$ GRAVITY WITH VARIABLE DECELERATION PARAMETER}}

\author{P.K. SAHOO, PARBATI SAHOO}

\address{Department of Mathematics,\\ Birla Institute of Technology and Science-Pilani,\\
Hyderabad Campus,\\
Hyderabad-500078, India\,\\
\email{pksahoo@hyderabad.bits-pilani.ac.in, sahooparbati1990@gmail.com} }

%

\author{BINAYA K. BISHI}

\address{Department of Mathematics,\\ Visvesvaraya National Institute of Technology,\\ Nagpur-440010,
India.\\
binaybc@gmail.com}

\maketitle

\begin{history}
\received{(06 Feb 2017)}
\revised{(23 Feb 2017)}
\end{history}

\begin{abstract}
The objective of this work enclosed with the study of spatially homogeneous anisotropic Bianchi type-I universe in $f(R,T)$ gravity (where $R$ is the Ricci scalar and $T$ is the trace of stress energy momentum tensor) in two different cases viz. $f(R,T)=R+2f(T)$ and $f(R,T)=f_1(R)+f_2(T)$ with bulk viscosity matter content. In this study, we consider a time varying deceleration parameter, which generates an accelerating universe to obtain the exact solution of the field equations. The physical and kinematical properties of both the models are discussed in detail for the future evolution of the universe. We have explored the nature of WEC, DEC, SEC and energy density for both the cases. We have found that both the models, with bulk viscosity matter component, show an acceleration of the universe. We have also shown that the cosmic jerk parameter is compatible with the three kinematical data sets.
\end{abstract}

\section{Introduction}	
The revolution in the understanding of the present universe has become possible due to the research in observational cosmology in the past two decade. Recent observational data from Plank collaboration \cite{Plank15}, Baryon Oscillation Spectroscopic Survey (BOSS) \cite{BOSS16} and Atacama Cosmology Telescope Polarimeter (ACTPol) collaboration \cite{ACTPol14} provides relevant experimental evidence about the acceleration of our universe. In the last century, modern cosmology reached a new vision to establish considerable advancements in the account of the current accelerated expanding universe. The two crucial observational groups including Supernovae cosmology project and the high-redshift Supernovae search team have provided the main evidence for the cosmic acceleration of the universe \cite{Garnavich998,Riess98,Perlmutter99,Garnavich98,Perlmutter97,Letelier83}. The other cosmic observations like cosmic microwave background (CMB) fluctuations \cite{Spergel03,Spergel07}, large-scale structure (LSS) \cite{Tegmark04,Daniel08}, cosmic microwave radiation (CMBR) \cite{Caldwell004,Huang06} indicate that the present universe is undergoing an accelerated expansion. It is also believed that the universe changed with time from early deceleration phase to late time acceleration phase \cite{Caldwell06}. Now, a team of scientists led by Professor Subir Sarkar \cite{Nielsen16} has raised the question on this standard cosmological concept about the acceleration, which is also supported by Shariff et al. \cite{Shariff16}.\\
The two promising approaches confirmed by the cosmological research community for discussing the cosmic expansion of the universe are: \newline
The first one is the introduction of the most exotic entity of mysterious universe dubbed as dark energy, which has positive energy density and negative pressure. Recently, from plank cosmological results and Wilkson microwave anisotropic probe (WAMP) 9 years results \cite{Planck13}, it may be concluded that the universe embodied with 68.5\% dark energy, 26.5\% of dark matter and 5\% of baryonic matter. Dark energy can be represented either in terms of cosmological constant or by the help of equation of state parameter (EOS) $\omega=\frac{p}{\rho}$, where $p$ is the pressure and $\rho$ is the energy density.\\
The second approach to picturize the evolution of the universe is a modified version of Einstein's field equations of general relativity, which can be established through the Einstein-Hilbert action principle. In this process, the matter Lagrangian is replaced by an arbitrary function. Thereafter, these modified theories become most attractive aspirant to observe the accelerated expansion of the universe as well as the effective causes related to dark energy.\\
In addition, it has been postulated that the standard Einstein-Hilbert action is modified by an arbitrary function $f(R)$, where $R$ is Ricci scalar curvature. The $f(R)$ gravity become an adequate theory to provide the gravitational alternative for dark energy and about the early inflation plus late-time cosmic acceleration of the universe \cite{Capozziello02,Caroll004,Dolgov03,Nojiri03,Nojiri04,Abdalaa05,Mena06,Bamba08}. In 2007, the $f(R)$ gravity theory is restructured by merging the matter Lagrangian density $L_m$ with initial arbitrary function of the Ricci scalar $R$ \cite{Betrolami07}. The unification of dark energy and early time inflation with late time acceleration from $f(R)$ theory to all Lorentz non-invariant theories is discussed by Nojiri and Odintsov \cite{Nojiri11}. Through continuation of this work of coupling, in 2011, Harko et al. \cite{Harko11} proposed a new modified theory named as $f(R,T)$ theory, where the gravitational part of the action still depends on the Ricci scalar $R$ like $f(R)$ theories and also a function of trace $T$. It is suggested that due to the matter-energy coupling, the leading model of this theory depends on source term representing the variation of energy-momentum tensor. Indefinitely many modified gravitational theories such as $f(G)$ gravity, $f(R,G)$ gravity, and $f(T)$ gravity etc. were developed to achieve the accelerated expansion of the Universe. Myrzakulov et al. \cite{Myrzakulov15} have investigated the inflation in $f(R,\phi)$ theories of gravity where the scalar field is coupled with gravity. Sebastiani and Myrzakulov \cite{Sebastiani15} have briefly reviewed various $f(R)$ gravity models for inflation, in particular, Starobinsky-like inflation. After that, $f(R,T)$ gravity become most prominent theory for investigating the fate of the late time accelerating expansion of the universe. A phase transition also occurred from matter-dominated era to an accelerated phase during the reconstruction of $f(R,T)$ gravity theory \cite{Houndjo12}. In the context of common perfect fluid matter, an axially symmetric cosmological model was constructed in the framework of $f(R,T)$ gravity \cite{Sahoo014}. In $f(R,T)$ gravity theory, many cosmological models can be constructed changing choices in matter source. Sahoo and Mishra \cite{Sahoo14} constructed Kaluza-Klein dark energy model by considering the wet dark fluid matter source. Recently, Moraes et al. \cite{Moraes16} derived the Starobinksy model in $f(R,T)$ gravity.\\
By considering the metric dependent Lagrangian density $L_m$, the respective field equation for $f(R,T)$ gravity is formulated from the Hilbert-Einstein variational principle in the following manner:
\begin{equation}
S=\int \sqrt{-g}\biggl(\frac{1}{16\pi G}f(R,T)+L_{m}\biggr)d^{4}x
\end{equation}%
where, $L_{m}$ is the usual matter Lagrangian density of matter source, $f(R,T)$ is an arbitrary function of
Ricci scalar $R$ and the trace $T$ of the energy-momentum tensor $T_{ij}$ of
the matter source, and $g$ is the determinant of the metric tensor $g_{ij}$. The energy-momentum tensor $T_{ij}$ from Lagrangian matter is defined in the form
\begin{equation}
T_{ij}=-\frac{2}{\sqrt{-g}}\frac{\delta (\sqrt{-g}L_{m})}{\delta g^{ij}}
\end{equation}%
and its trace is $T=g^{ij}T_{ij}$.\newline
Here, we have assumed that the matter Lagrangian $L_{m}$ depends only on the
metric tensor component $g_{ij}$ rather than its derivatives. Hence, we
obtain
\begin{equation}
T_{ij}=g_{ij}L_{m}-\frac{\partial L_{m}}{\partial g^{ij}}
\end{equation}%
By varying the action $S$ in Eq. (1) with respect to $g_{ij}$, the $f(R,T)$
gravity field equations are obtained as
\begin{multline}
f_{R}(R,T)R_{ij}-\frac{1}{2}f(R,T)g_{ij}+(g_{ij}\Box -\nabla _{i}\nabla
_{j})f_{R}(R,T)\\=8\pi T_{ij}-f_{T}(R,T)T_{ij}-f_{T}(R,T)\Theta _{ij}
\end{multline}%
where,
\begin{equation}
\Theta _{ij}=-2T_{ij}+g_{ij}L_{m}-2g^{lm}\frac{\partial ^{2}L_{m}}{\partial
g^{ij}\partial g^{lm}}
\end{equation}
Here, $f_{R}(R,T)=\frac{\partial f(R,T)}{\partial R}$, $f_{T}(R,T)=\frac{%
\partial f(R,T)}{\partial T}$, $\Box \equiv \nabla ^{i}\nabla _{i}$ where $%
\nabla _{i}$ is the covariant derivative.\newline
Contracting Eq. (4), we get
\begin{equation}
f_{R}(R,T)R+3\Box f_{R}(R,T)-2f(R,T)=(8\pi -f_{T}(R,T))T-f_{T}(R,T)\Theta
\end{equation}%
where $\Theta =g^{ij}\Theta _{ij}$.\newline
From Eqs (4) and (6), the $f(R,T)$ gravity field equations takes the form
\begin{multline}
f_{R}(R,T)\biggl(R_{ij}-\frac{1}{3}Rg_{ij}\biggr)+\frac{1}{6}f(R,T)g_{ij}= \\
8\pi -f_{T}(R,T)\biggl(T_{ij}-\frac{1}{3}Tg_{ij}\biggr)-f_{T}(R,T)\biggl(%
\Theta _{ij}-\frac{1}{3}\Theta g_{ij}\biggr)+\nabla _{i}\nabla _{j}f_{R}(R,T)
\end{multline}%
It is worth mentioning here that the physical nature of the matter field through $%
\Theta _{ij}$ is used to form the field equations of $f(R,T)$ gravity. To construct different kinds of cosmological models according to the choice of matter source, Harko et al. \cite{Harko11} constructed
three types of $f(R,T)$ gravity as presented below
\begin{equation}
f(R,T)=\left\{
\begin{array}{lcl}
R+2f(T) &  &  \\
f_{1}(R)+f_{2}(T) &  &  \\
f_{1}(R)+f_{2}(R)f_{3}(T) &  &
\end{array}%
\right.
\end{equation}%
The individual field equations for each frames of $f(R,T)$ gravity is given as \\
Case-I: $f(R,T)=R+2f(T)$
\begin{equation}
R_{ij}-\frac{1}{2}Rg_{ij}=8\pi T_{ij}-2f'(T)T_{ij}-2f'(T)\Theta_{ij}+f(T)g_{ij}
\end{equation}
Case-II: $f(R,T)=f_1(R)+f_2(T)$
\begin{multline}
f_1'(R)R_{ij}-\frac{1}{2}f_1(R)g_{ij}+(g_{ij}\Box -\nabla _{i}\nabla
_{j})f_1(R)\\=8\pi T_{ij}-2f_2'(T)T_{ij}-2f_2'(T)\Theta_{ij}+\frac{1}{2} f_2(T)g_{ij}
\end{multline}
Most of the researchers constructed cosmological models in the presence of perfect fluid matter to analyse the accelerated expansion of the universe. Recent observations provide the evidence for the accelerated expansion of the universe due to the presence of an unknown form of energy (dark energy) having negative pressure. On that account, we need to construct a cosmological model for expanding universe without considering dark energy and dark matter, while choosing most reliable matter component. It is believed that the cosmic viscosity acts as the dark energy candidate which may play an important role in causing the accelerated expansion of the universe by consuming negative effective pressure \cite{Zimdahl01}. At present, there is great interest in formulating the cosmological model with dissipative fluid  matter rather than with dust (a pressure-less distribution) or with a perfect fluid matter, which gives more realistic model instead of others and most effective to pay attention to the dynamical background of homogeneous and isotropic universe. It is commonly accepted that, in the early phase of the universe, the matter behaved like a viscous fluid during the neutrino decoupling in radiation era \cite{Misner67,Misner68,Klimerk76}. The nature of singularity  occurred for perfect fluid can be modified through the dissipative mechanism of viscous fluid. This viscous fluid cosmological model helps to explain the matter distribution on the large entropy per baryon in the present universe. Also, the phase transition and string creation are involved with viscous effects as per the Grand Unified Theories (GUT). It has been found that the mixture of minimally coupled self-interacting scalar field can successfully derive an accelerated expansion of the universe, while the same mixture with perfect fluid unable to do it \cite{Chimento02}. Hence, many cosmological models with viscous fluid in the early universe have been widely discussed in the literature \cite{Singh11,Yadav12,Jamil12,Singh07,Hu06}. \newline
Again, observations have been conducted to obtain the homogeneity and isotropic properties of the universe. It is believed that at the end of the inflationary era, the geometry of the universe was homogeneous and isotropic \cite{Linde08}, where the FLRW models played an important role is representing both spatially homogeneous and isotropic universe. But the theoretical argument and the anomalies found in CMB provide the evidence for the existence of an anisotropic phase, which is later called isotropic one. After the announcement of Plank probe results \cite{Ade14}, it is  believed that the early universe may not have been exactly uniform. Thus, the existence of inhomogeneous and anisotropic properties of the universe has gained popularity when it comes to constructing cosmological models under the supervision of anisotropic background. Therefore, Bianchi type models are very relevant for describing the early universe with the anisotropic background. Due to some analytical difficulties in studying the inhomogeneous models, many researchers considered the Bianchi type models for investigating the cosmic evolution of the early universe as they are homogeneous and anisotropic. There exist nine types ($I-IX$) of Bianchi space-times in literature.  Here, we consider Bianchi type-I space-time, as it is the simplest spatially homogeneous and anisotropic. It is also known as the immediate generalization of the FLRW flat metric with different scale factors in each spatial direction. In some special cases, the Bianchi type-I models include \textit{Kasner metric}, which helps to govern the dynamics near the singularity. The Bianchi type-I cosmological models are more compatible with the simplest mathematical form which attracts various researcher to study different aspects. The nature of Bianchi type-I cosmological model has been studied in the context of a viscous fluid. It has been observed that viscosity can cause the qualitative behaviour of solutions near the singularity without removing the total initial big bang singularity \cite{Belinskii76}. Recently, the bulk viscous matter content has been discussed along with cosmological constant in the framework of Bianchi type-I space time \cite{Bali08}. Later, the spatially homogeneous anisotropic Bianchi type-I model with bulk viscous fluid matter was studied with the assumption of constant deceleration parameter in the framework of $f(R,T)$ gravity theory through the explanation of the energy conditions \cite{Singh09,Fabris06,Saha07,Sharif012,Sharif 013,Sharif14}.\newline
This work is motivated by the aforesaid literature. The study is presented in different sections. Section-I deals with the basic formalism of $f(R,T)$ gravity field equations from Einstein-Hilbert action along with some essential literature review. In section-II, we derive the exact solution of both the cases of $f(R,T)$ gravity ($f(R,T)=R+2f(T)$ and $f(R,T)=f_1(R)+f_2(T)$) for the spatially homogeneous anisotropic Bianchi type-I space-time with the help of time varying deceleration parameter. The energy density, bulk viscous pressure, bulk viscous coefficient, the trace of matter, Ricci scalar, the energy conditions are derived and their graphical representations are also discussed in this section. In section-III, physical parameters of both the models are presented. The detail discussion of the figures are covered in section-IV. The last section contains the conclusion of the covered study.
\section{Field equations and Solutions}
We consider the spatially homogeneous LRS Bianchi type-I metric as
\begin{equation}
ds^{2}=dt^{2}-A^{2}dx^{2}-B^2(dy^{2}+dz^{2})
\end{equation}
where $A, B$ are functions of cosmic time $t$ only.\\
The energy momentum tensor for bulk viscus fluid is considered in the form
\begin{equation}
T_{ij}=(\rho+\overline{p})u_{i}u_{j}-\overline{p}g_{ij}
\end{equation}
where $u^i=(0,0,0,1)$ is the four velocity vector in co-moving coordinate system satisfying $u_iu_j=1,$
\begin{equation}
\overline{p}=p-3\xi H
\end{equation}
is the bulk viscous pressure which satisfies the linear equation of state $p=\gamma \rho, \ \ 0\leq \gamma \leq 1$, $\xi$ is the bulk viscous coefficient, $H$ is Hubble's parameter, $p$ is pressure and $\rho$ is the energy density.\\
The trace of energy momentum tensor is given as
\begin{equation}
T=\rho-3\overline{p}
\end{equation}
\subsection{Case-I: $f(R,T)=R+2f(T)$}
The field equation (9) with $f(T)=\alpha T$, where $\alpha$ is an arbitrary constant for the metric (11) is obtained as
\begin{eqnarray}
-2\frac{\ddot{B}}{B}-\frac{\dot{B}^{2}}{B^{2}}=(8\pi+3\alpha)\overline{p}-\alpha \rho  \\
-\frac{\ddot{A}}{A}-\frac{\ddot{B}}{B}-\frac{\dot{A}\dot{B}}{AB}=(8\pi+3\alpha)\overline{p}-\alpha \rho \\
2\frac{\dot{A}\dot{B}}{AB}+\frac{\dot{B}^{2}}{B^{2}}=(8+3\alpha) \rho-\alpha \overline{p}
\end{eqnarray}
Dot represent derivatives with respect to time $t.$\\
The deceleration parameter (DP) is defined as
\begin{equation}
q=-\frac{a\ddot{a}}{\dot{a}^2}
\end{equation}
where, $a$ is the average scale factor.\\
We have three equations (15-17) involving four parameters as $A, B, \overline{p} \ \&\ \rho$.
In order to solve these equations, we assume the time varying deceleration parameter as
\begin{equation}
q=-1+\frac{\beta}{1+a^{\beta}}
\end{equation}
where, $\beta>0$ is a constant.\\
The Hubble's parameter is defined as $H=\frac{\dot{a}}{a}$, and from above equation we obtain
\begin{equation}
H=\frac{\dot{a}}{a} =1+a^{-\beta}
\end{equation}
where, the integrating constant is assumed as unity. Integrating (20) we have found
\begin{equation}
a= (e^{\beta t}-1)^{\frac{1}{\beta}}
\end{equation}
Setting $a(t)=\frac{1}{(1+z)}$, where $z$ is the redshift, leads to relation
\begin{equation}
t=\frac{\log \left(\left(\frac{1}{z+1}\right)^B+1\right)}{B}
\end{equation}
The corresponding $q(z)$ is obtained as
\begin{equation}
q=\frac{B}{\left(\frac{1}{z+1}\right)^B+1}-1
\end{equation}
The deceleration parameter describes the evolution of the universe . The cosmological models of the evolving universe transits from early decelerating phase ($q>0$) to current accelerating phase ($q<0$). Whereas, the models can be classified on the basis of the time dependence of DP. Recent observations like SNe Ia \cite{Riess98} and CMB anisotropy \cite{Bennet03} confirmed that the present universe is undergoing  an accelerated phase of expansion and the value lies in between $-1\leq q\leq 0$. Fig-1 depicts the behavior of deceleration parameter with respect to redshift, in which the value of $q$ lies in specified range of accelerating phase. The transition phase of deceleration parameter from deceleration to acceleration occurs at some transition redshift $z_{tr}$ which is completely decided by the parameter $\beta$. In this work, we have considered three representative values of $\beta$ i.e. 1.4772, 1.5 and 1.55 corresponding to $z_{tr} =0.65$, 0.5874 and
0.4706 respectively. The values of transition redshift $z_{tr}$ for our model are agreeing with the observational data \cite{Capozziello14,Capozziello15,Farooq17}. The transition from deceleration to acceleration phase in $f(R,T)$ gravity with polynomial function of $T$ is discussed by Moraes et al. \cite{Moraes16}. Our first case  clearly shows that the model is completely under accelerated phase which is  conformity with observational data. 

\begin{figure}[h!]
\centering
\includegraphics[width=75mm]{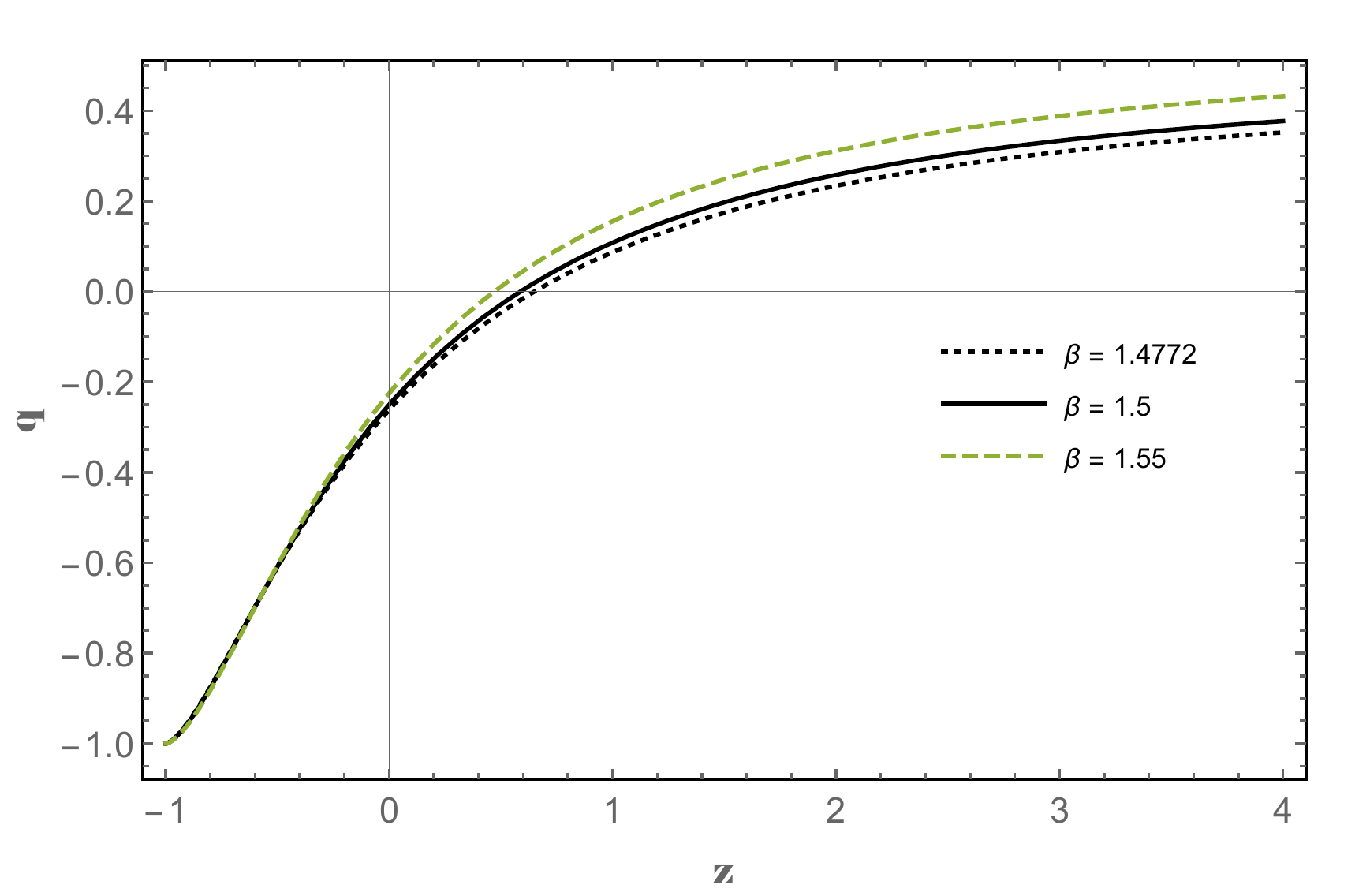}
\caption{$q$  vs. redshift with different $\beta$.}
\end{figure}

The volume is defined as $V=a^3=AB^2$. Using (21) in this relation, the values of the metric potentials $A, B$ are obtained as
\begin{eqnarray}
A= (e^{\beta t}-1)^{\frac{2}{\beta}} \\
B= (e^{\beta t}-1)^{\frac{1}{2 \beta}}
\end{eqnarray}
Consequently metric (11) takes the form
\begin{equation}
ds^{2}=dt^{2}-(e^{\beta t}-1)^{\frac{4}{\beta}}dx^{2}-(e^{\beta t}-1)^{\frac{1}{\beta}}(dy^{2}+dz^{2})
\end{equation}
Solving the field equations (15-17), the values of $\rho$ and $\overline{p}$ are obtained as
\begin{equation}
\rho=\frac{1}{(8\pi+3\alpha)^2-\alpha^2}\biggl[\biggl(18\pi+\frac{3\alpha}{2}+\frac{5\alpha\beta}{2}\biggr)e^{2\beta t} (e^{\beta t}-1)^{-2}-\frac{5\alpha \beta}{2}e^{\beta t} (e^{\beta t}-1)^{-1}\biggr]
\end{equation}
\begin{multline}
\overline{p}= \frac{-1}{(8 \pi +3\alpha)^2-\alpha^2}\biggl[\biggl(42\pi+\frac{27\alpha}{2}-\frac{5(8\pi+3\alpha)\beta}{2}\biggr)e^{2\beta t} (e^{\beta t}-1)^{-2}\\+\frac{5(8\pi+3\alpha) \beta}{2}e^{\beta t} (e^{\beta t}-1)^{-1}\biggr]
\end{multline}
We can observe that the energy density remains positive throughout the evolution of the universe and is a decreasing function of cosmic time $t$. It starts with a positive value and approaches to zero as $t\rightarrow \infty$. The bulk viscous pressure $\overline{p}$ is an increasing function of time, which begins from a large negative value and tends to zero at present epoch. As per the observation, the negative pressure is due to DE in the context of accelerated expansion of the universe. Hence, the behavior of bulk viscous pressure in our model is agreed with this observation.\\
The coefficient of bulk viscosity $\xi$ and the pressure are obtained as
\begin{multline}
\xi=\frac{1}{(8 \pi +3\alpha)^2-\alpha^2}\biggl[\biggl(2\pi(3\gamma+7)+\frac{\alpha(\gamma+9)}{2}-\frac{5\beta(\alpha\gamma-8\pi-3\alpha)}{6}\biggr)e^{\beta t} (e^{\beta t}-1)^{-1}\\-\frac{5\beta(\alpha\gamma-8\pi-3\alpha)}{6}\biggr]
\end{multline}
\begin{equation}
p=\gamma\rho
=\frac{\gamma}{(8\pi+3\alpha)^2-\alpha^2}\biggl[\biggl(18\pi+\frac{3\alpha}{2}+\frac{5\alpha\beta}{2}\biggr)e^{2\beta t} (e^{\beta t}-1)^{-2}-\frac{5\alpha \beta}{2}e^{\beta t} (e^{\beta t}-1)^{-1}\biggr]
\end{equation}

\begin{figure}[h!]
\centering
\minipage{0.50\textwidth}
  \includegraphics[width=65mm]{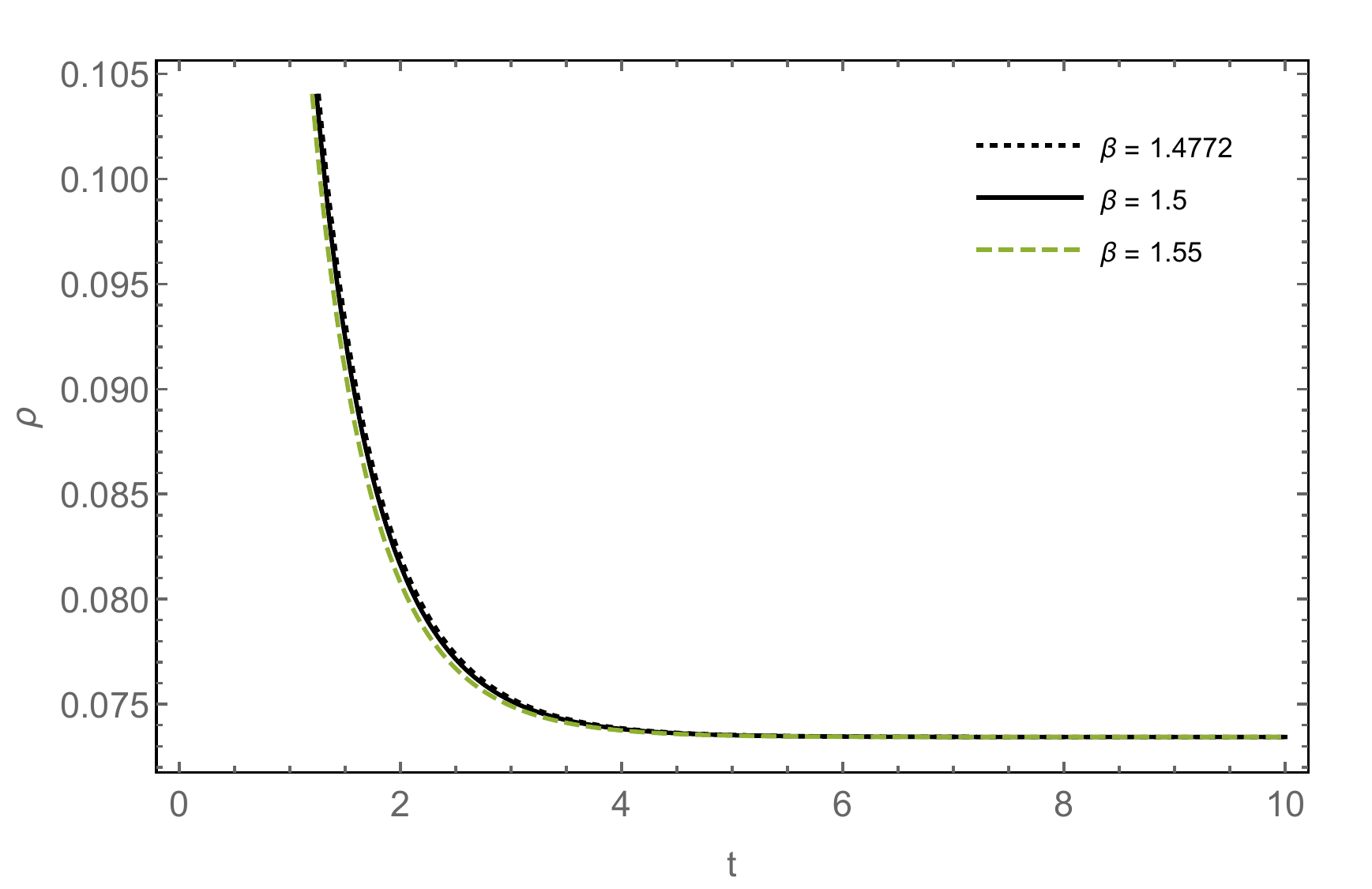}
  \caption{$\rho$  vs. time with $\alpha=1$ and different $\beta$.}
\endminipage\hfill
\minipage{0.50\textwidth}
  \includegraphics[width=65mm]{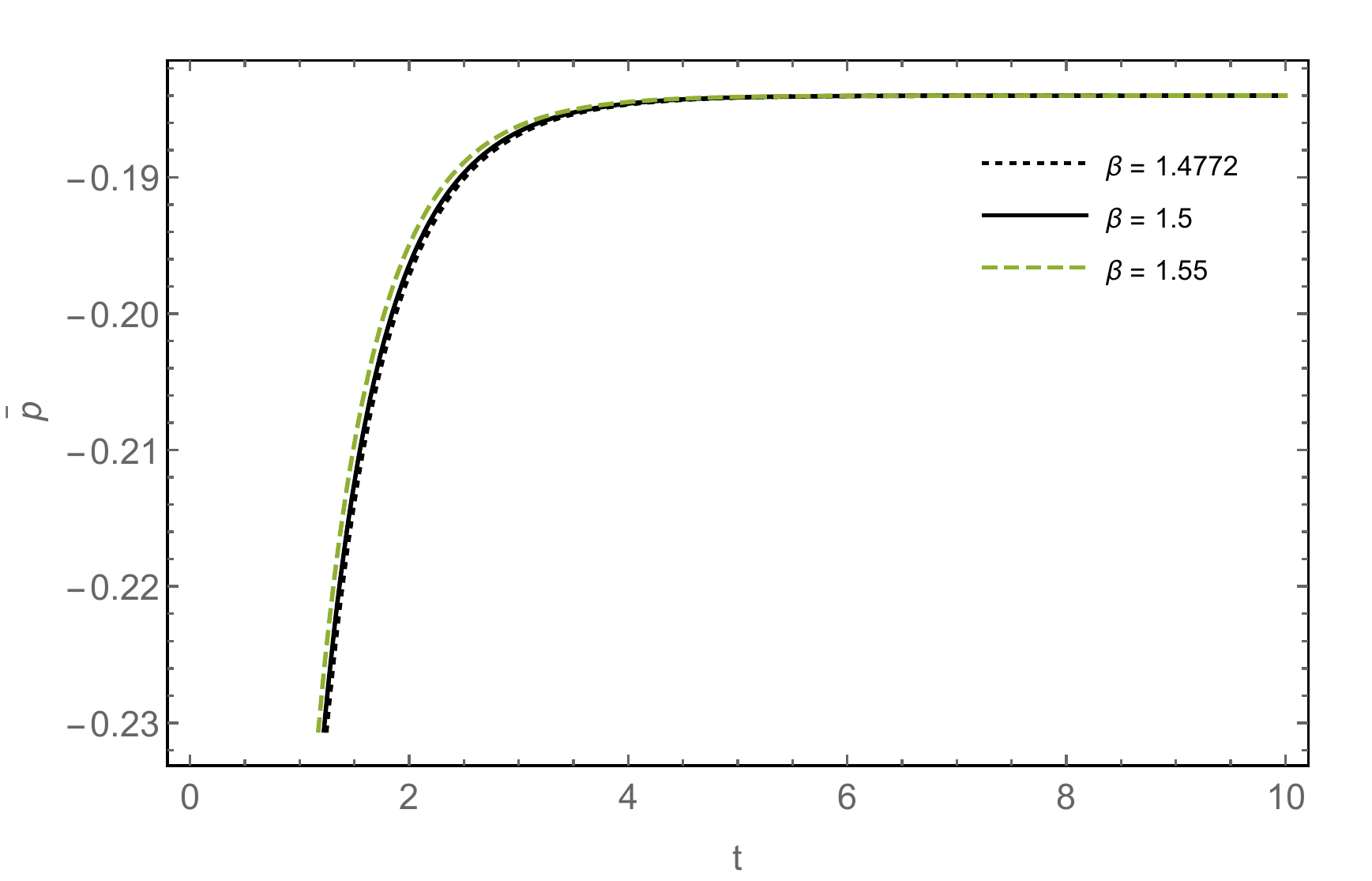}
  \caption{$\overline{p}$  vs. time with $\alpha=1$ and different $\beta$.}
\endminipage\hfill
\minipage{0.70\textwidth}%
  \includegraphics[width=65mm]{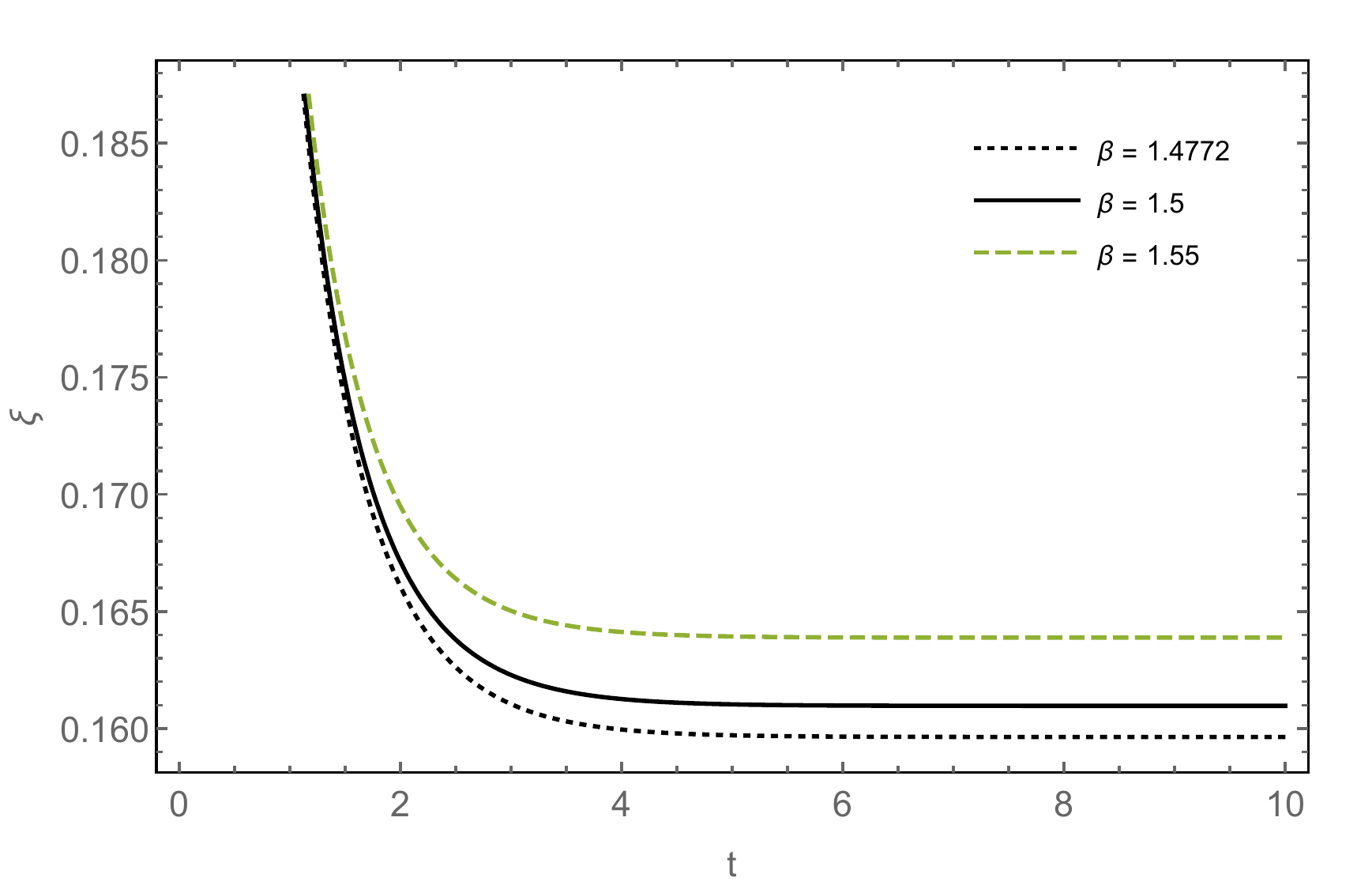}
  \caption{$\xi$  vs. time with $\alpha=1,\gamma=0.5$ and different $\beta$.}
\endminipage
\end{figure}

Figure 4, indicates that the bulk viscous coefficient $\xi$ is positive through out the universe and becomes finite as $t\rightarrow \infty$ for our model.\\
The energy conditions are some alternative conditions for matter content of the theory. In general relativity (GR), the role of these energy conditions is to prove  the theorems about the existence of space-time singularity and black holes  \cite{Wald84}.
The energy conditions are used in many approaches to understand the evolution of the universe. Here, we discussed some of the most popular energy conditions for this model. The weak energy conditions (WEC), dominant energy conditions (DEC) and strong energy conditions (SEC) are given as
\begin{eqnarray}
\rho>0, \ \ \rho-p\geqslant 0 \ \ (WEC) \\
\rho+p\geqslant 0 \ \ (DEC)  \\
\rho+3p\geqslant 0 \ \ (SEC)
\end{eqnarray}
These are used in various aspects through their own importance. For example, the DEC gives an idea about the stability of matter source and imposes the dark energy along with equation of state parameter $\omega$ for lower bound $\omega\geq -1$, which may cause Big Rip \cite{Carroll03}. SEC violation is a typical trait of a positive cosmological constant $\Lambda$ \cite{Lake04}. Finally, WEC shows that the matter-energy is always non-negative. In $f(R,T)$ gravity, the matter support of wormholes is examined by the behavior of energy conditions and the nature of solution for FRW model with perfect fluid matter is studied through energy conditions \cite{Zubair16,Sharif13}. According to the predefined literature, one can consider the energy conditions to be useful to analyze the behavior of cosmological solutions throughout the universe. Therefore, we dealt with some well-known energy conditions like WEC, DEC and SEC to observe our solutions in both of the cases. Figures(5-7) show the behavior of WEC, DEC, and SEC  with the proper choice of constants respectively.\\

\begin{figure}[h!]
\centering
\minipage{0.48\textwidth}
\includegraphics[width=65mm]{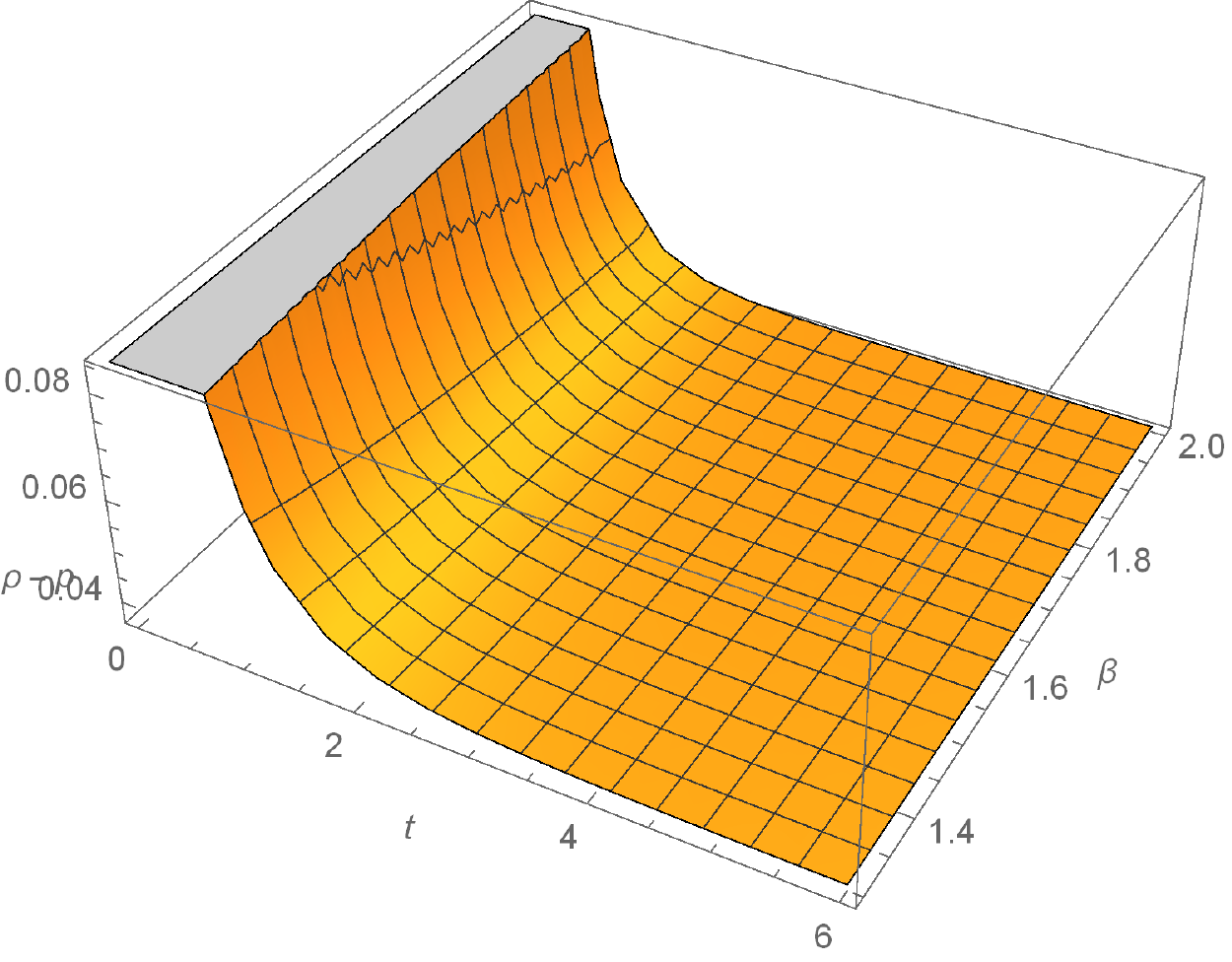}
  \caption{Behaviour of WEC versus $t$ and $\beta$ with $\alpha=1, \gamma=0.5 $.}
\endminipage\hfill
\minipage{0.48\textwidth}
  \includegraphics[width=65mm]{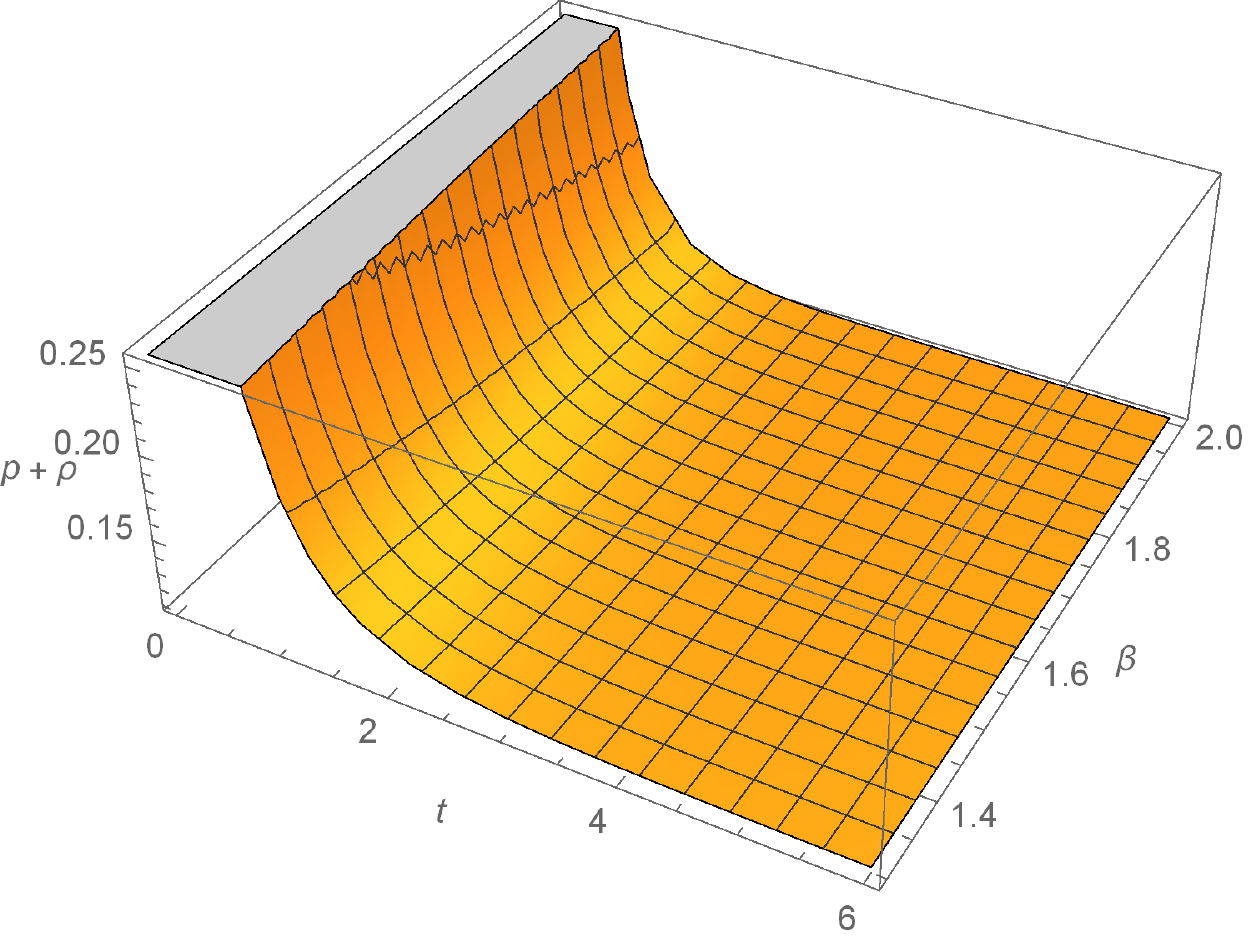}
  \caption{Behaviour of DEC versus $t$ and $\beta$ with $\alpha=1, \gamma=0.5 $.}
\endminipage\hfill
\minipage{0.70\textwidth}%
  \includegraphics[width=65mm]{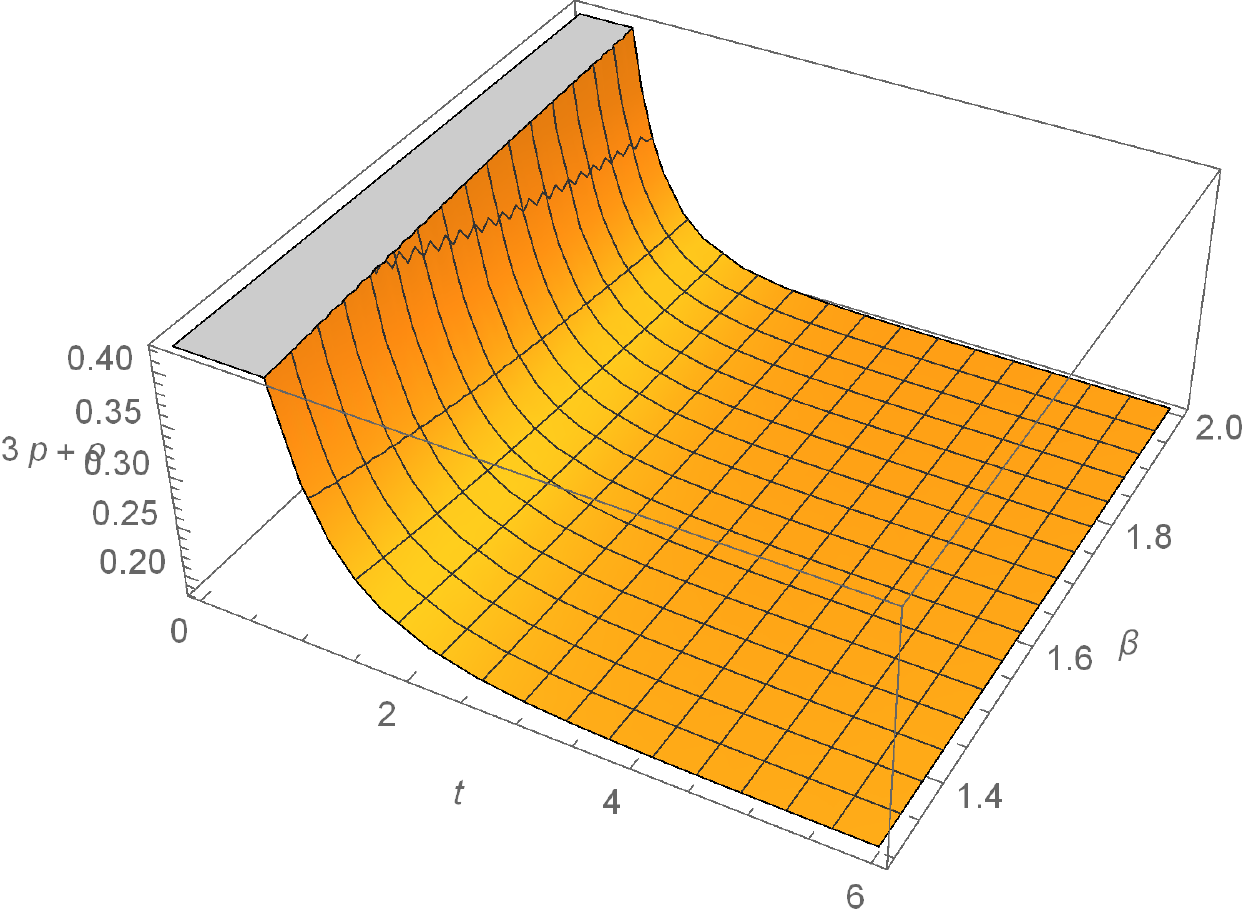}
  \caption{Behaviour of SEC versus $t$ and $\beta$ with $\alpha=1, \gamma=0.5$.}
\endminipage
\end{figure}

From figures 5-7, it is observed that all the energy conditions are satisfied for this model.\\
The values of Ricci scalar $R$ and the trace of matter source $T$ are obtained as
\begin{equation}
R=-\biggl[\frac{2\ddot{A}}{A}+\frac{4\ddot{B}}{B}+\frac{4\dot{A}\dot{B}}{AB}+\frac{\dot{B}^2}{B^2}\biggr] =\biggl(6\beta-\frac{27}{2}\biggr)e^{2\beta t}(e^{\beta t}-1)^{-2}-6\beta e^{\beta t}(e^{\beta t}-1)^{-1}
\end{equation}
\begin{multline}
T=\rho-3\overline{p}=\frac{1}{(8\pi+3\alpha)^2-\alpha^2}\biggl[(144\pi+42\alpha-(60\pi+20\alpha)\beta)e^{2\beta t}(e^{\beta t}-1)^{-2}\\+(60\pi+20\alpha)\beta e^{\beta t}(e^{\beta t}-1)^{-1}\biggr]
\end{multline}
Using the above equations, the function $f(R,T)$ is obtained as
\begin{multline}
f(R,T)=\biggl(6\beta-\frac{27}{2}+\frac{(288-120\beta)\alpha \pi+48\alpha^2-40\alpha^2 \beta}{(8\pi+3\alpha)^2-\alpha^2}\biggr)e^{2\beta t}(e^{\beta t}-1)^{-2}\\+\biggl(\frac{120\pi \alpha+40\alpha^2\beta}{(8\pi+3\alpha)^2-\alpha^2}-6\beta \biggr)e^{\beta t}(e^{\beta t}-1)^{-1}
\end{multline}

\begin{figure}[ht]
\centering
\includegraphics[width=75mm]{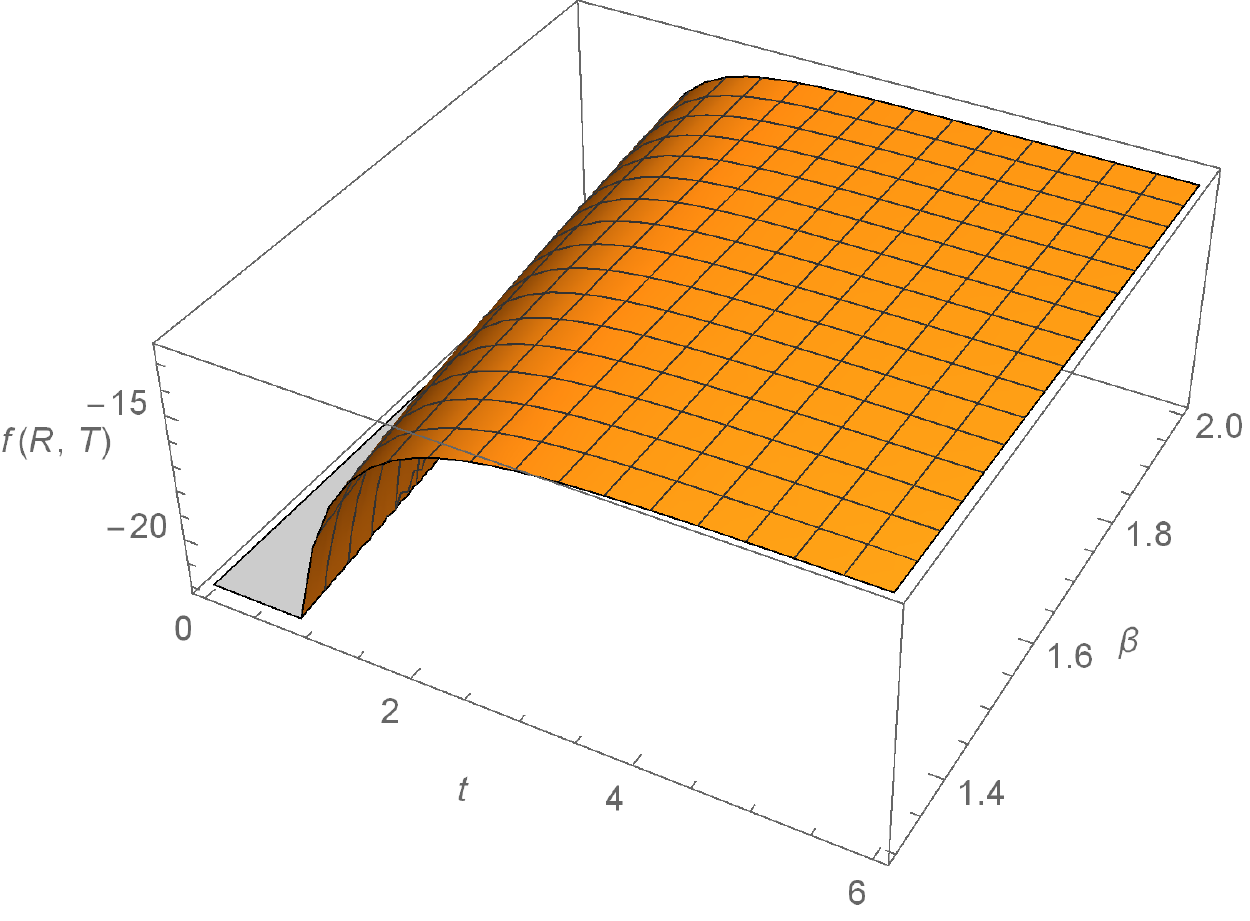}
\caption{Behaviour of $f(R,T)$ versus $t$ and $\beta$ with $\alpha=1$.}
\end{figure}
Fig. 8 shows the behaviour of the function $f(R,T)$ for this model.
\subsection{Case-II: $f(R,T)=f_1(R)+f_2(T)$}
In this case we assumed $f_1(R)=\mu R$ and $f_2(T)=\mu T$, where $\mu$ is an arbitrary constant. The corresponding field equations in the general form is
\begin{equation}
 R_{ij}-\frac{1}{2}Rg_{ij}=\biggl(\frac{8\pi +\mu}{\mu}\biggr)T_{ij}+\biggl(p+\frac{1}{2}T\biggr)g_{ij}=\chi T_{ij}+\biggl(p+\frac{1}{2}T\biggr)g_{ij}
\end{equation}
where $\chi=\biggl(\frac{8\pi +\mu}{\mu}\biggr)$. The set of field equations for the metric (11) are
\begin{eqnarray}
-2\frac{\ddot{B}}{B}-\frac{\dot{B}^{2}}{B^{2}}=\biggl(\chi+\frac{1}{2}\biggr)\overline{p}-\frac{1}{2} \rho  \\
-\frac{\ddot{A}}{A}-\frac{\ddot{B}}{B}-\frac{\dot{A}\dot{B}}{AB}=\biggl(\chi+\frac{1}{2}\biggr)\overline{p}-\frac{1}{2}\rho \\
2\frac{\dot{A}\dot{B}}{AB}+\frac{\dot{B}^{2}}{B^{2}}=\biggl(\chi+\frac{1}{2}\biggr)\rho-\frac{1}{2}\overline{p}
\end{eqnarray}
The above set of field equations (38-40) admits the same solutions (24) and (25) as obtained in the previous case.
Using the metric potentials from (24) and (25), the values of energy density $\rho$ and bulk viscous pressure $\overline{p}$ are
\begin{equation}
\rho=\frac{1}{\biggl(\chi+\frac{1}{2}\biggr)^2-\frac{1}{4}}\biggl[\biggl(\frac{9\chi-6}{4}+\frac{5\beta}{4}\biggr)e^{2\beta t} (e^{\beta t}-1)^{-2}-\frac{5\beta}{4}e^{\beta t} (e^{\beta t}-1)^{-1}\biggr]
\end{equation}
\begin{multline}
\overline{p}= \frac{-1}{\biggl(\chi+\frac{1}{2}\biggr)^2-\frac{1}{4}}\biggl[\biggl(\frac{21\chi+6}{4}-\frac{5\beta}{2}\biggl(\chi+\frac{1}{2}\biggr)\biggr)e^{2\beta t} (e^{\beta t}-1)^{-2}\\+\frac{5\beta}{2}\biggl(\chi+\frac{1}{2}\biggr)e^{\beta t} (e^{\beta t}-1)^{-1}\biggr]
\end{multline}
Fig. 9 shows that the energy density $\rho$ is decreasing function of time and remains always positive. It converges to zero as $t\rightarrow \infty$. Fig. 10 depicts the variation of $\overline{p}$ versus cosmic time $t$. It is observed from the figure that bulk viscous pressure is increasing function of time from a large negative value and is approaching to zero at present time. \\
For this case, the values of bulk viscosity coefficient $\xi$ and effective pressure $p$ are
\begin{multline}
\xi=\frac{\gamma \rho-\overline{p}}{3H}
= \frac{1}{\biggl(\chi+\frac{1}{2}\biggr)^2-\frac{1}{4}}\biggl[\biggl(\frac{9\gamma+21-10\beta}{12}\chi+\frac{(5\beta-6)(\gamma-1)}{12}\biggr)e^{\beta t} (e^{\beta t}-1)^{-1}\\-\frac{5\gamma \beta}{12}+\frac{5\beta}{6}\biggl(\chi+\frac{1}{2}\biggr)\biggr]
\end{multline}
\begin{equation}
 p =\gamma\rho=\frac{\gamma}{\biggl(\chi+\frac{1}{2}\biggr)^2-\frac{1}{4}}\biggl[\biggl(\frac{9\chi-6}{4}+\frac{5\beta}{4}\biggr)e^{2\beta t} (e^{\beta t}-1)^{-2}-\frac{5\beta}{4}e^{\beta t} (e^{\beta t}-1)^{-1}\biggr]
\end{equation}

\begin{figure}[ht]
\centering
\minipage{0.50\textwidth}
  \includegraphics[width=65mm]{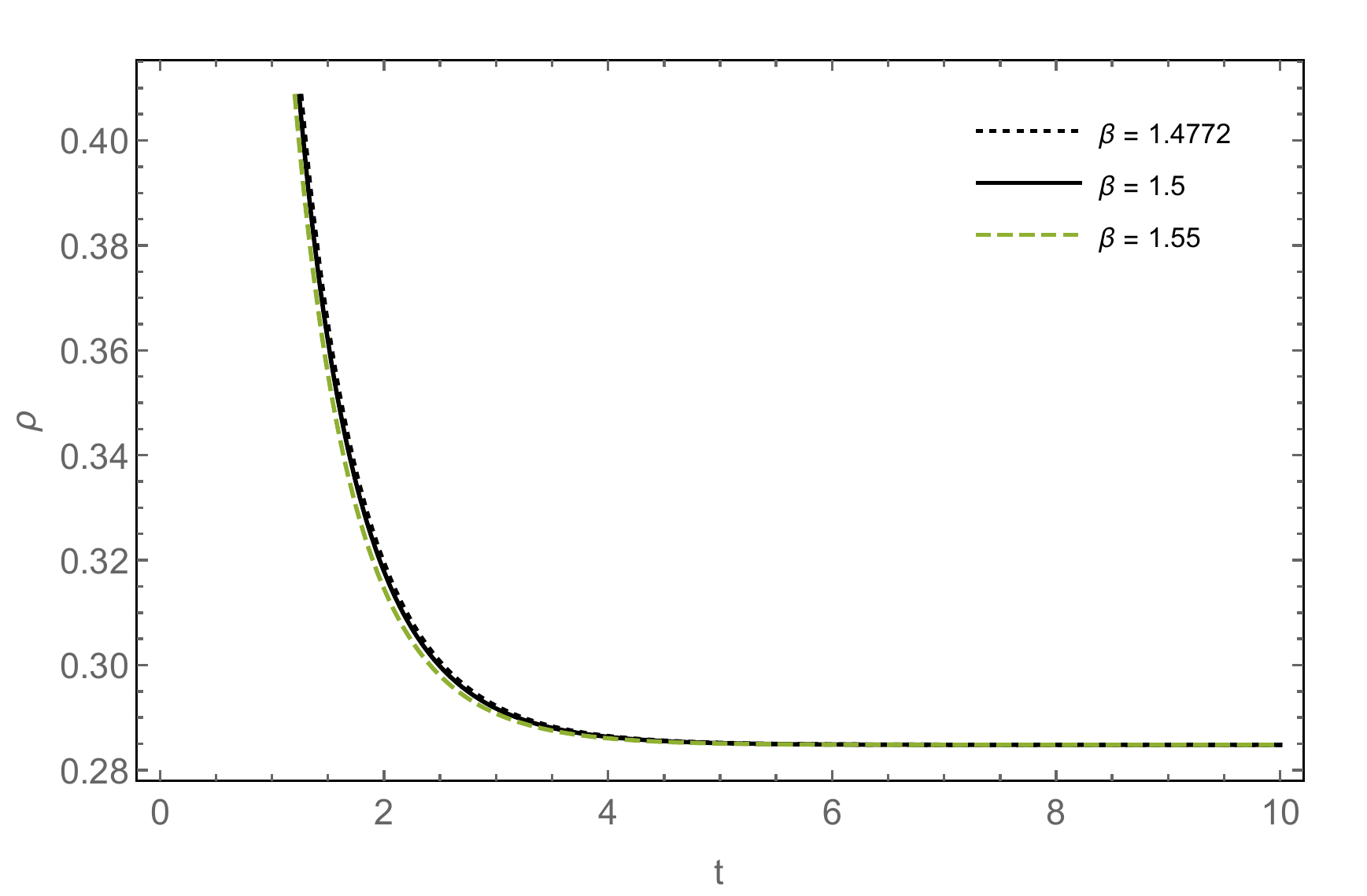}
  \caption{$\rho$  vs. time with  $\mu=5$ and different $\beta$.}
\endminipage\hfill
\minipage{0.50\textwidth}
  \includegraphics[width=65mm]{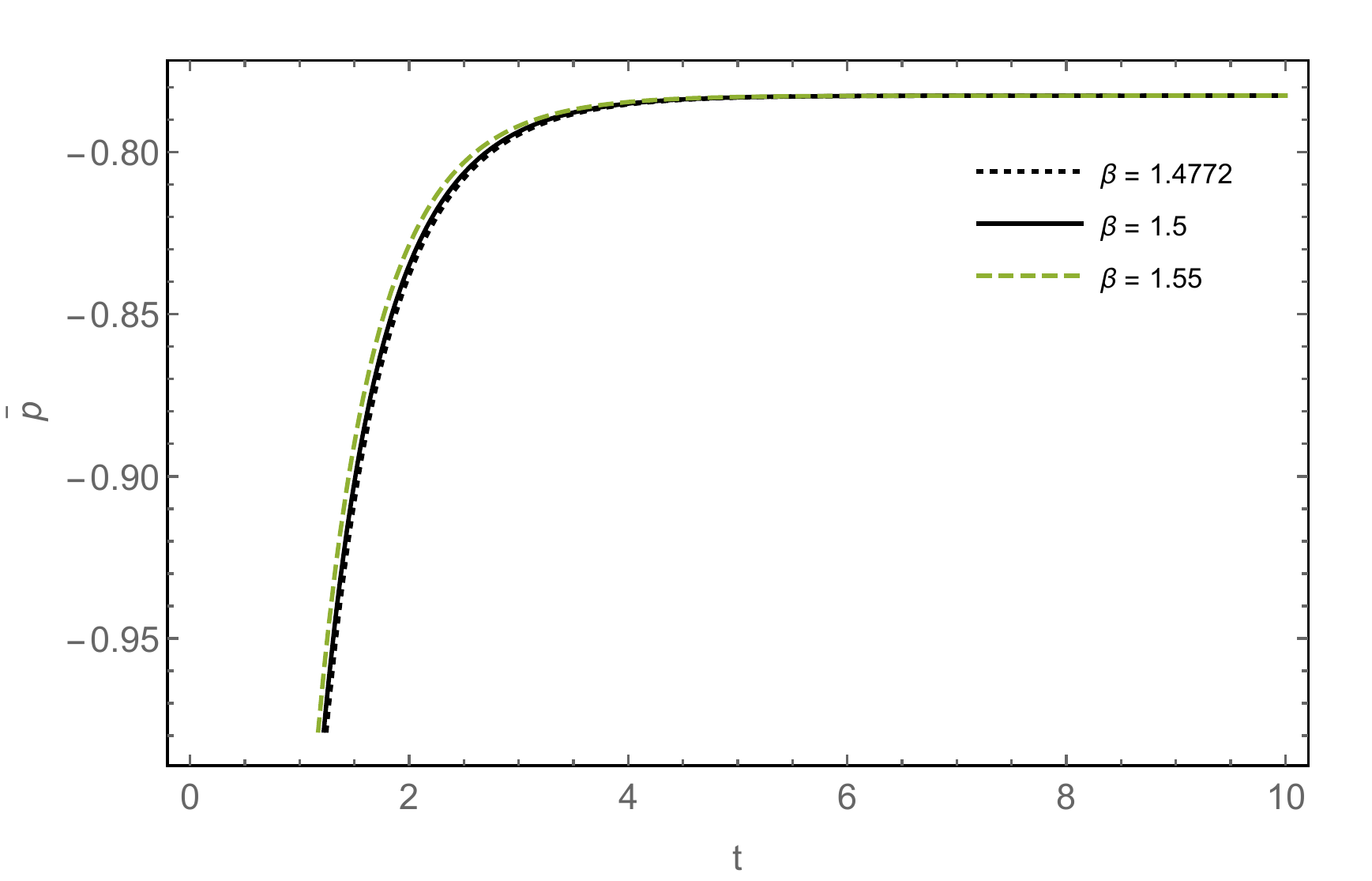}
  \caption{$\overline{p}$  vs. time with $\mu=5$ and different $\beta$.}
\endminipage\hfill
\minipage{0.75\textwidth}%
  \includegraphics[width=65mm]{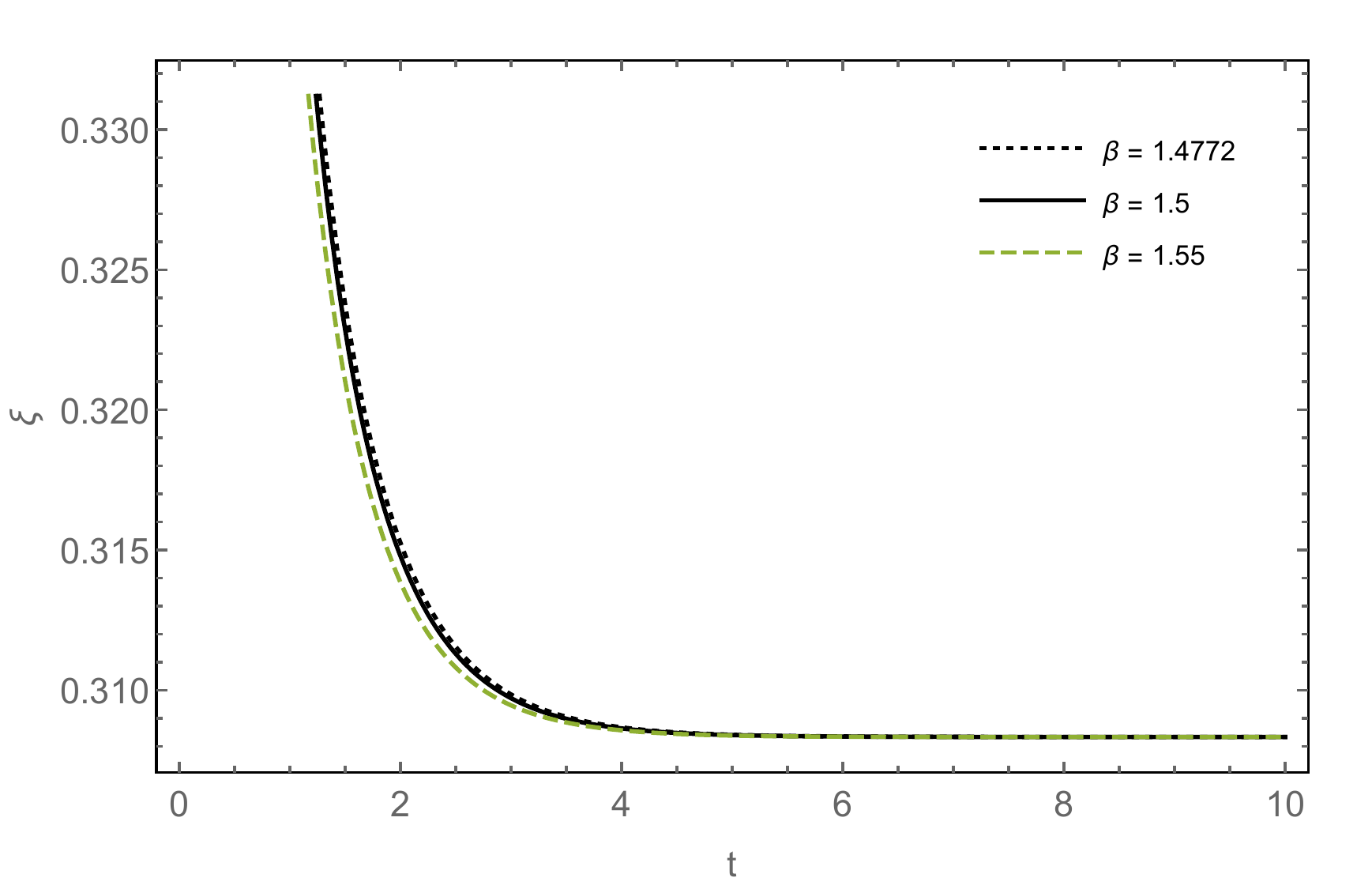}
  \caption{$\xi$  vs. time with $\mu=5$ and different $\beta$}
\endminipage
\end{figure}

As suggested in figure 11, the bulk viscosity coefficient is constant throughout the universe as required.\\
The weak energy conditions (WEC), dominant energy conditions (DEC) and strong energy conditions (SEC) for this model are plotted below.\\

\begin{figure}[ht]
\centering
\minipage{0.48\textwidth}
  \includegraphics[width=65mm]{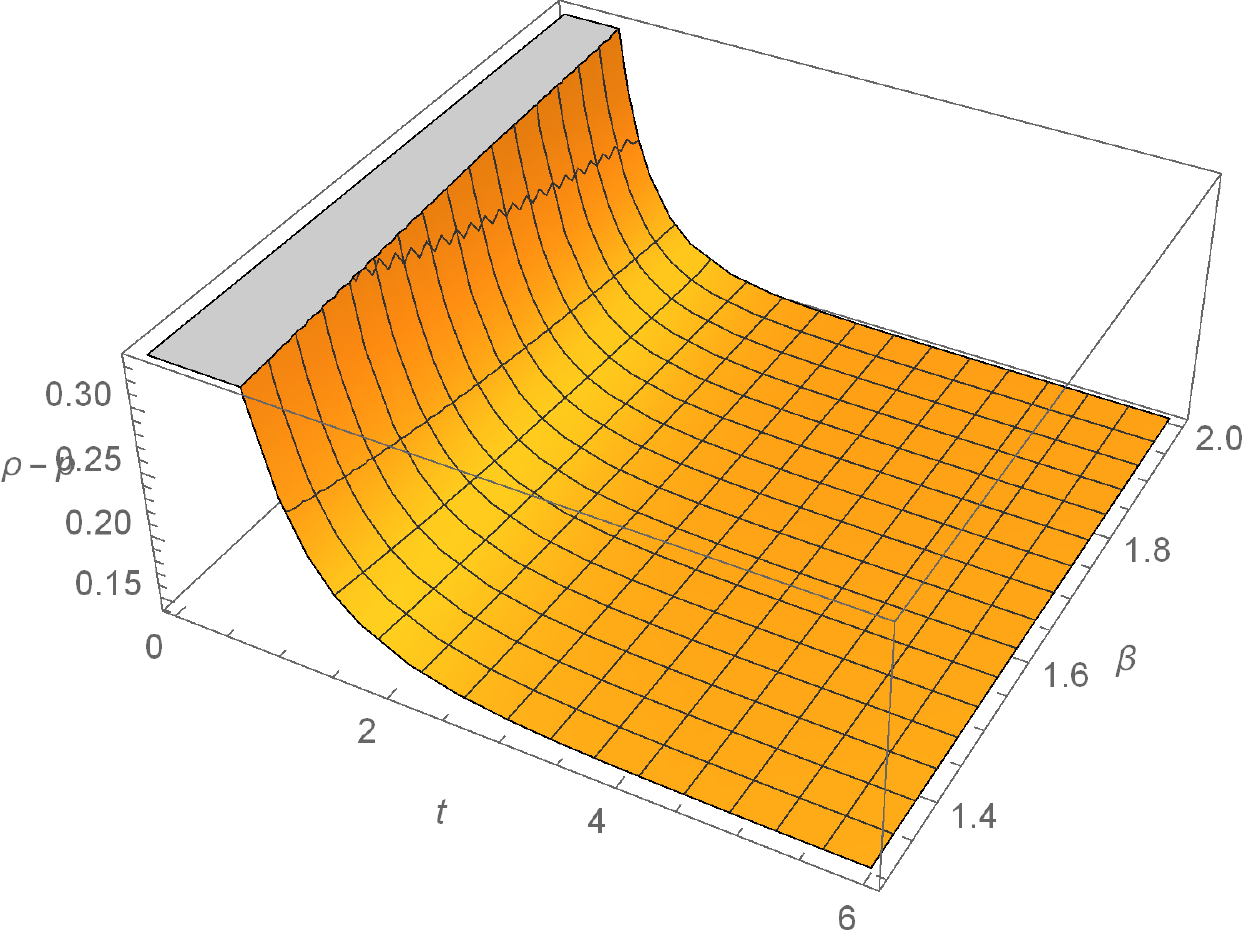}
  \caption{Behaviour of WEC versus $t$ and $\beta$  with  $\mu=5$.}
\endminipage\hfill
\minipage{0.48\textwidth}
  \includegraphics[width=65mm]{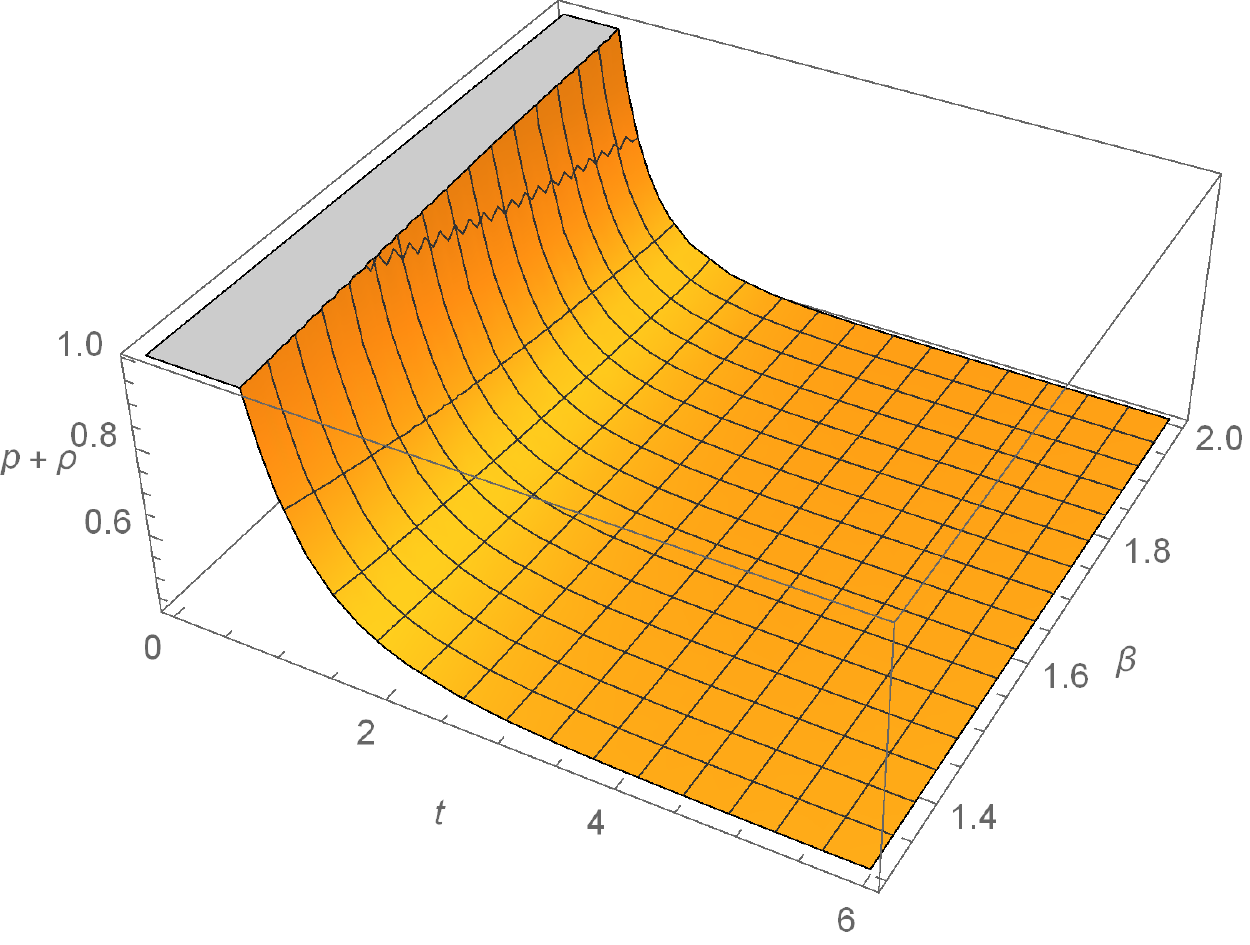}
  \caption{Behaviour of DEC versus $t$ and $\beta$ with $\mu=5$.}
\endminipage\hfill
\minipage{0.70\textwidth}%
  \includegraphics[width=65mm]{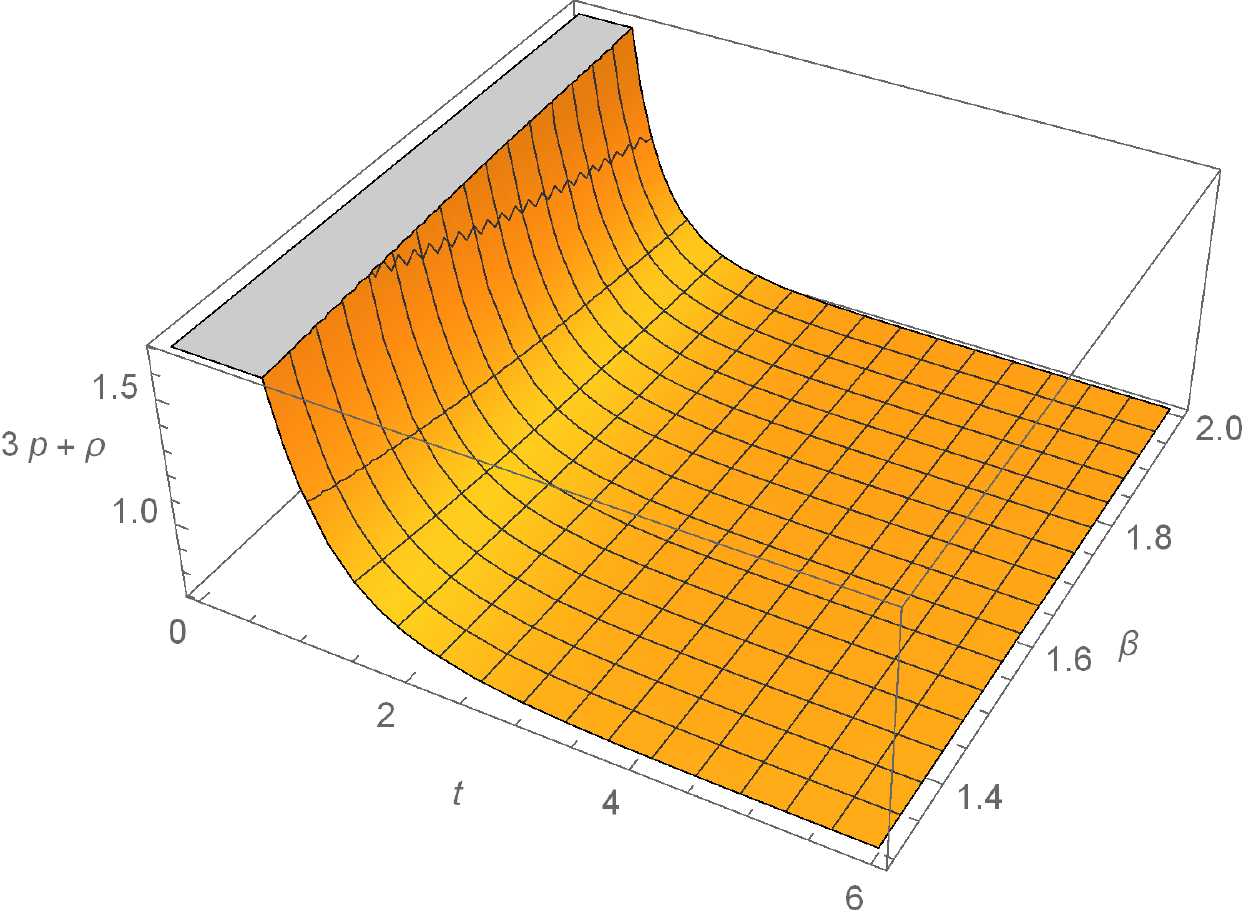}
  \caption{Behaviour of SEC versus $t$ and $\beta$ with $\mu=5$.}
\endminipage
\end{figure}

Figures 12-14, show that the energy conditions are completely agreed with GR.\\
We can obtain the trace of matter $T$ for this model as
\begin{multline}
T=\rho-3\overline{p}=\frac{1}{(\chi+\frac{1}{2})^2-\frac{1}{4}}\biggl[\biggl(\biggl(\frac{36-15\beta}{2}\biggr)\chi-\frac{12-10\beta}{4}\biggr)e^{\beta t}(e^{\beta t}-1)^{-2}\\+\biggl(\frac{15\beta \chi}{2}-\frac{10\beta}{4}\biggr)\beta e^{\beta t}(e^{\beta t}-1)^{-1}\biggr]
\end{multline}
The relation $f(R,T)$ for the above case is obtained in the form
\begin{multline}
f(R,T)=\biggl(6\beta-13.5+\frac{(18-7.5\beta)\chi-2.5\beta+3}{(\chi+\frac{1}{2})^2-\frac{1}{4}}\biggr)\mu e^{2\beta t}(e^{ \beta t}-1)^{-2}\\+\biggl(\frac{7.5\beta \chi-2.5\beta}{(\chi+\frac{1}{2})^2-\frac{1}{4}}-6\beta \biggr)\mu e^{\beta t}(e^{\beta t}-1)^{-1}
\end{multline}

\begin{figure}[ht]
\centering
\includegraphics[width=75mm]{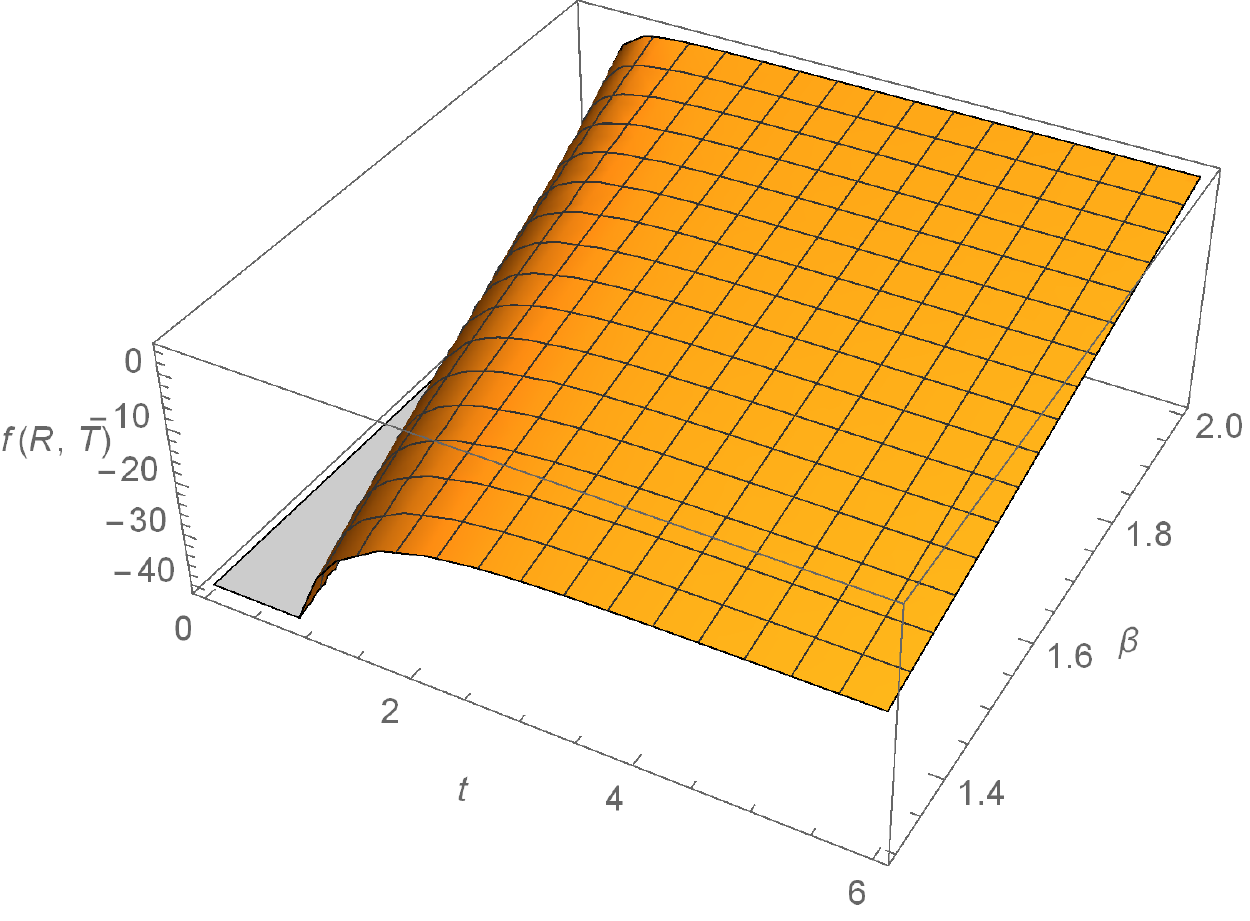}
\caption{Behaviour of $f(R,T)$ versus $t$ and $\beta$ with  $\mu=5$.}
\end{figure}
Fig. 15 shows the behaviour of $f(R,T)$ for $f(R,T)=f_1(R)+f_2(T)$ model.
\section{Physical properties of the models}
The rate of expansion of the universe with respect to time is defined by Hubble's parameter as well as deceleration parameter. For detail kinematical descriptions of the cosmological expansions can be obtained by taking in account of some extended set of parameters having higher order time derivatives of the scale factor.\\
The spatial volume turns out to be
\begin{equation}
V=AB^2=(e^{\beta t}-1)^{\frac{3}{\beta}}
\end{equation}
The above equation indicates that in both the models the spatial volume is zero at initial time $t=0$. It shows that the evolution of our universe starts with big bang scenario. It is further noted that from (21) the average scale factor becomes zero at  the initial epoch. Hence, both models have a point type singularity \cite{Maclum71}. The spatial volume increases with time.\\
The Hubble's parameter $H$, expansion scalar $\theta$ and shear scalar $\sigma^2$ become
\begin{equation}
H=\frac{1}{3}(H_1+2H_2)=e^{\beta t} (e^{\beta t}-1)^{-1}
\end{equation}
\begin{equation}
\theta=3H=3e^{\beta t} (e^{\beta t}-1)^{-1}
\end{equation}
\begin{equation}
\sigma^2=\frac{1}{2}\biggl(H_1^2+2H_2^2-\frac{\theta^2}{3}\biggr)= \frac{3}{4} e^{2\beta t} (e^{\beta t}-1)^{-2}
\end{equation}
From the above equations, we can observe that the Hubble factor, scalar expansion and shear scalar diverge at $t=0$ and they become finite as $t\rightarrow \infty$. It is noted here that the isotropic condition $\frac{\sigma^2}{\theta^2}$ becomes constant (from early to late time), which shows that the model does not approach isotropy throughout the evolution of the universe. The anisotropy parameter
\begin{equation}
\Delta =\frac{1}{3}\sum_{i=1}^{3}\biggl(\frac{H_{i}-H}{H}\biggr)^{2}=6\biggl(
\frac{\sigma }{\theta }\biggr)^{2}=\frac{1}{2}
\end{equation}
The anisotropic parameter becomes constant for our models. From the above mentioned equation it can be observed that our models are expanding and accelerating universe which starts at a big bang singularity.\\
\textbf{Jerk parameter:}\\
The jerk parameter is considered as one of the important quantity for describing the dynamics of the universe. The models close to $\Lambda$CDM can be described by the cosmic jerk parameter $j$ \cite{Sahni02, Visser05}. For flat $\Lambda$CDM model the value of jerk is $j=1$ \cite{Rapetti06}.  Jerk parameter is a dimensionless third derivative of scale factor $a$ with respect to cosmic time $t$ and is defined as
\begin{equation}
j=\frac{a^2}{\dot{a}^3}\frac{d^3 a}{dt^3}
\end{equation}
The above expression can be written in terms of deceleration parameter as
\begin{equation}
j=q+2q^2-\frac{\dot{q}}{H}
\end{equation}
Thus, the jerk parameter for our models is
\begin{equation}
j=1-3\beta e^{-\beta t}+2\beta^2 e^{-2\beta t}+\beta^2e^{-2\beta t}(e^{\beta t}-1)
\end{equation}

\begin{figure}[ht]
\centering
\includegraphics[width=75mm]{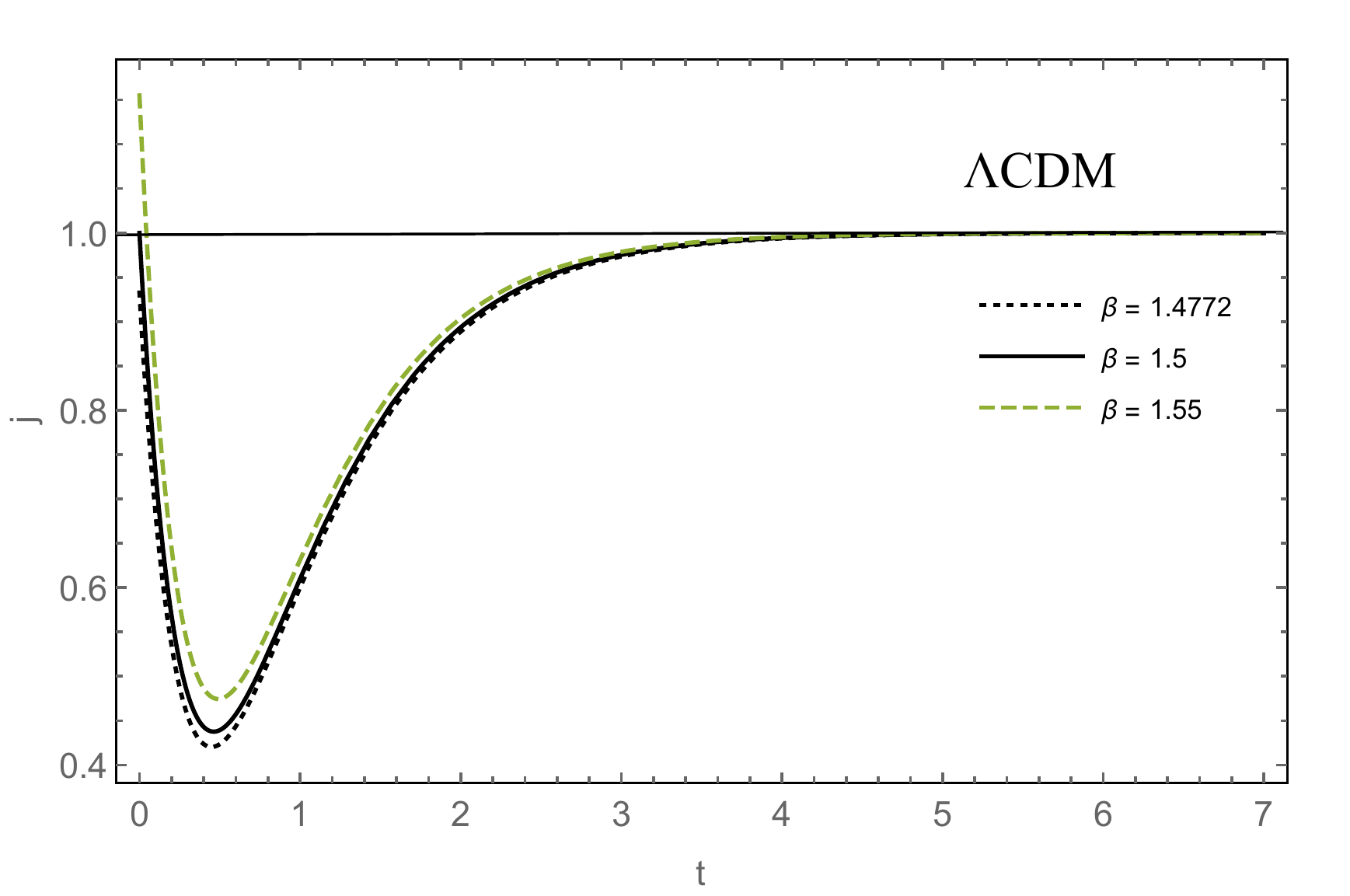}
\caption{Behaviour of Jerk parameter versus $t$ with different $\beta$.}
\end{figure}

From the Fig. 16, it is clear that our value does not overlap with the value $j=2.16^{+0.81}_{-0.75}$ obtained from combination of three kinematical data sets: the gold sample data of type Ia supernovae \cite{Riess04}, the SNIa data obtained from the SNLS project \cite{Astier06}, and the X-ray galaxy cluster distance measurements \cite{Rapetti06}. We have plotted the jerk parameter for different values of $\beta$ in Fig. 16. One can observe that the jerk parameter remains positive through out the universe and  is equal to the $\Lambda$CDM model at $t\geq5.5$ for the considered values of $\beta$.  It is interesting to note that our model is close to  $\Lambda$CDM model for the following set of values as presented in table-I.\\
\textbf{$r-s$ parameter:}\\
The state-finder pair $\{r,s\}$ is defined as \cite{Sahni03}
\begin{equation}
r=\frac{\dddot{a}}{aH^3}, \ \ \ \ s=\frac{r-1}{3(q-\frac{1}{2})}
\end{equation}
The state-finder pair is a geometrical diagnostic parameter, which is constructed from a space-time metric directly, and it is more universal compared to physical variables, which depend on the properties of physical fields describing DE,
since physical variables are model dependent. For the flat $\Lambda$CDM model the state-finder pair obtained as $\{r,s\}=\{1,0\}$ \cite{Feng08}. The values of the state-finder parameter for our model are
\begin{equation}
r=1-3\beta e^{-\beta t}+2\beta^2 e^{-2\beta t}+\beta^2e^{-2\beta t}(e^{\beta t}-1)
\end{equation}
\begin{equation}
s=\frac{1}{6 \beta-9 e^{\beta t}}\biggl[2\beta^2 e^{-\beta t}(e^{\beta t}-1)+4\beta^2 e^{-\beta t}-6\beta\biggr]
\end{equation}
From the expressions of $r$ and $s$ parameters, we found that $\{r,s\}=\{1,0\}$ only when $t=\frac{1}{\beta}\ln\bigl(\frac{\beta}{3-\beta}\bigr)$. The variation of $\beta$ and $t$ for $\{r,s\}=\{1,0\}$  is presented in table-I. For the set of values of $(\beta,t)$ our models represents $\Lambda$CDM models, which are presented in table-I. The $r-s$ trajectory of our models is presented in the Fig. 17.
\begin{table}[ht]
\begin{tabular}{|c|c|c|c|}
  \hline
  $\beta$ & $t=\frac{1}{\beta}\ln\bigl(\frac{\beta}{3-\beta}\bigr)$ & $r$  &  $s$\\\hline
  1.5 & 0 & 1 & $\infty$ \\
  1.6 & 0.08345 & 1 & $-1.458333324\times 10^{-9}\approx 0$ \\
  1.7 & 0.15780 & 1 & 0 \\
  1.8 & 0.22525 & 1 & 0 \\
  1.9& 0.28765 & $1.000000002\approx 1$ & 0 \\
  2 & 0.5$\ln(2)$ & 1 & 0\\
  \hline
\end{tabular}
\caption{Variation of $\beta$ and $t$ for $\{r,s\}$}
\end{table}
From the above table-I, it is observed that at initial epoch $t=0$ the parameter $r$ becomes unity and $s$ becomes finite and diverges for $\beta=1.5$.

\begin{figure}[ht]
\centering
\includegraphics[width=75mm]{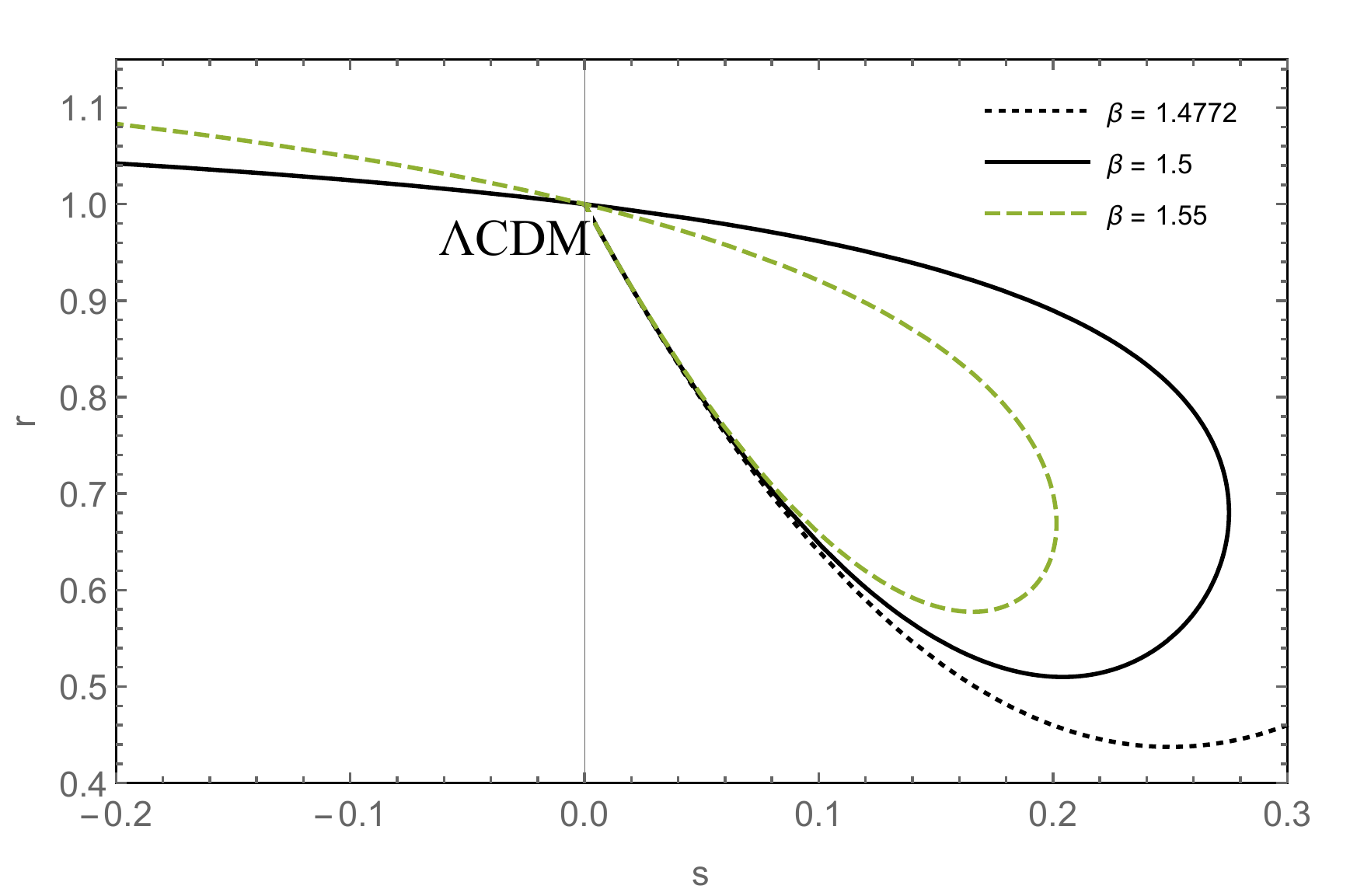}
\caption{$r$  vs. $s$.}
\end{figure}
\section{Discussion}
It is well known that the standard $\Lambda$CDM (cosmological constant $\Lambda$+Cold Dark Matter) model fits well to the currently available observational data. The most recent data from Planck+BAO+JLA+$H_0$ yields the matter density parameter $\Omega_m=0.3089$ in the $\Lambda$CDM model \cite{Plank15}. Using this value, we find that the $\Lambda$CDM universe transits from deceleration to acceleration at the redshift $z_{tr}=0.65$. In our models, it corresponds to $\beta=1.4772$ as may be seen in Figure 1, where we show the variation of deceleration parameter vs redshift for $\beta=1.4772$. Thus the models present here have a transition from deceleration to acceleration with transition redshift $z_{tr}$ satisfying the observational data. Figure 2 and Figure 9 depict the variation of energy density against time in presence of different parameters as presented in the figure for case-I and case-II respectively. The figures indicate that, in both the cases energy density is positive valued and decreasing function of time. It is also approaching zero with the evolution of time.
The variation of bulk viscous pressure and coefficient of bulk viscosity is presented in the Figure 3 and Figure 10 and Figure 4 and Figure 11 for case-I and case-II respectively. From the figures, we noticed that bulk viscous pressure and coefficient of bulk viscosity are negative and positive valued function of time respectively. Also, the coefficient of bulk viscosity decreases with the evolution of time and maintains a constant rate after $t>4$. The variation of energy conditions against time for both the cases is presented from Figure 5 to Figure 7 and Figure 12 to   Figure 14 respectively. In both the cases, energy conditions (WEC, SEC, DEC) are satisfied. All the physical parameters presented in both the cases follow the same quantitative behaviour as that of observational data.
\section{Conclusion}
In this article, we have investigated the LRS Bianchi type I cosmological model in presence of bulk viscosity in the framework of $f(R,T)$ gravity. According to the choice of $f(R,T)$ we have presented two cosmological models.  The exact solutions of the modified Einstein's field equations are obtained under the choice of deceleration parameter of the form (19). The observations of both the models are as follows:
\begin{itemize}
  \item Both the models presented here are accelerating and the expanding Universe models follow an exponential expansion.
  \item Energy density and coefficient of bulk viscosity are positive valued and decreasing function of time in both the cases and also $\rho \rightarrow 0$ when $t\rightarrow\infty$.
  \item Bulk viscosity pressure $(\bar{p})$ is negative valued in both the cases.
  \item Energy conditions (SEC, WEC, DEC) are satisfied for both cases.
  \item Jerk parameter and state-finder trajectory in the $r-s$ plane are close to $\Lambda$CDM model.
\end{itemize}
\section{Acknowledgements}
We are very indebted to the  editor and the anonymous referee for illuminating suggestions that have significantly
improved our paper in terms of  research  quality as well as presentation.

\end{document}